\documentclass[aps,prl,showpacs,twocolumn,superscriptaddress,floats,amssymb]{revtex4}
\usepackage{amsmath}
\usepackage{amssymb}
\usepackage{xcolor}

\definecolor{purple}{rgb}{1,0,1}
\definecolor{lightblue}{cmyk}{0.12,0,0,0}
\definecolor{lightred}{cmyk}{0,0.1,0,0}
\definecolor{lightyellow}{cmyk}{0,0,0.1,0}
\definecolor{grey}{cmyk}{0.01,0.01,0.01,0.03}
\definecolor{darkred}{cmyk}{0,0.8,0,0.5}
\definecolor{lightred}{cmyk}{0,0.1,0,0}
\definecolor{dGreen}{cmyk}{1,0,1,0.4}
\definecolor{DdGreen}{cmyk}{1,0,1,0.7}
\definecolor{lightGreen}{cmyk}{0.1,0,0.1,0.03}
\definecolor{lb}{cmyk}{0.05,0.1,0.15,0.02}

\definecolor{brown}{rgb}{0.45,0,0}
\definecolor{lightblue}{cmyk}{0.15,0,0,0}\definecolor{lightred}{cmyk}{0,0.1,0,0}
\definecolor{lightyellow}{cmyk}{0,0,0.1,0}

\def \nn{\nonumber}
\def \dsp{\displaystyle}
\def \sss{\scriptscriptstyle}

\def\Akstpi{\big|A_{{\sss\! K^{\!*}\!\!\pi}}\big|}

\def\Akstkno{\big|A_{{\sss\! K^{\!*}\!\!K}}\big|}

\def\Akpipi{\big|A_{{\sss\!K_{\!s}\!\pi\!\pi}}\big|} 
\def\kstpi{{\sss\! K^{\!*}\!\!\pi}}
\def\kstk{{\sss K^{\!*{\!+}}\!\!K^{\!-}}}
\def\kstkbar{{\sss K^{\!*\!-}\!\!K^{\!+}}}
\def\kstkno{{\sss K^{\!*}\!\!K}}
\def\kpi{{\sss K\!\pi}}
\def\kk{{\sss K\!K}}
\def\yp#1{{Y^{\sss (+)}_{\!#1}}}
\def\ym#1{{Y^{\sss (-)}_{\!#1}}}

\def\barp{{\raise.35ex\hbox{$~{\sss (}$}}--{\raise.35ex\hbox{${\sss )}$}}}
\def\Dbarp{\hbox{$D$\kern-1.2em\raise1.4ex\hbox{\barp}}}
\def\Bar#1{\hbox{$#1$\kern-1.65em\raise1.6ex\hbox{\barp}}~}
\def\Over#1{\hbox{$#1$\kern-0.7em\raise1.6ex\hbox{--}}\kern0.15em}
\def\bBar#1{\hbox{$#1$\kern-1.2em\raise1.4ex\hbox{{\raise.35ex\hbox{$~{\sss (}$}}--{\raise.35ex\hbox{${\sss )}$}}}}}

\begin{document}

\title{Accurate measurement of the $\mathbf{ D^0-\bar{D}^0}$ mixing parameters}

\author{Nita Sinha}
\author{Rahul Sinha}
\affiliation{The Institute of Mathematical Sciences, Taramani, Chennai
  600113, India}

\author{T. E. Browder}
\affiliation{Department of Physics and Astronomy, University of
  Hawaii, Honolulu, HI 96822, USA}

\author{N.~G.~Deshpande}
\affiliation{Institute of Theoretical Science, University of Oregon,
  Eugene, OR 94703, USA} 

\author{Sandip Pakvasa}
\affiliation{Department of Physics and Astronomy, University of
  Hawaii, Honolulu, HI 96822, USA}

\date{\today}

\begin{abstract}
  We propose a new method to determine the mass and width differences
  of the two $D$ meson mass-eigenstates as well as the CP violating
  parameters associated with $D^0-\bar{D}^0$ mixing.  We show that an
  accurate measurement of all the mixing parameters is possible for an
  arbitrary CP violating phase, by combining observables from a time
  dependent study of D decays to a doubly Cabibbo suppressed mode with
  information from a CP eigenstate.  As an example we consider $D^0\to
  K^{*0} \pi^0$ decays where the $K^{*0}$ is reconstructed in both
  $K^+\pi^-$ and $K_{\sss\!S}\pi^0$. We also show that decays to the
  CP eigenstate $D\to K^+K^-$ together with $D\to K^+\pi^-$ decays can
  be used to extract all the mixing parameters.  A combined analysis
  using $D^0\to K^{*0} \pi^0$ and $D\to K^+K^-$ can also be used to
  reduce the ambiguity in the determination of parameters.
\end{abstract}

\pacs{14.40.Lb, 11.30.Er, 13.25.Fc, 12.60.-i}
\maketitle

Evidence for mixing in the neutral $D$ meson system has recently been
reported~\cite{Aubert:2007wf, Staric:2007dt, Abe:2007rd} by the Belle
and BaBar collaborations.  These experiments find non-vanishing width
and mass differences between the two neutral $D$ mass eigenstates
assuming negligible CP violation.  In this letter we propose a method
to determine all the mixing parameters accurately allowing for
arbitrary CP violation.

Within the Standard Model CP violation in the $D$ system is
negligible.  Hence observation of CP violation would be a good signal
for New Physics (NP)~\cite{Blaylock:1995ay}.  While no CP violation
has been seen in $D-\bar{D}$ mixing~\cite{Abe:2007rd}, with the
current precision large possible NP contributions are not ruled out.

We show that using the doubly Cabibbo suppressed (DCS) mode $D\to
K^{*0}\pi^0$ and its conjugate modes, we can solve for all the
$D-\bar{D}$ mixing parameters.  This is possible if the
$K^{*0}/\bar{K}^{*0}$ is reconstructed both in the self tagging
$K^\pm\pi^\mp$ mode and in the CP eigenstate $K_{\sss\! S}\pi^0$ mode.
While the CP eigenstates $D\to K^+K^-$ cannot alone be used to
determine all the mixing parameters, we demonstrate that minimal
additional information from DCS modes allows determination of all
parameters. This approach may provide the optimal method to determine
all the parameters with current data.  In both these cases, the
parameters can be determined accurately even in the limit of a small
or vanishing CP violating mixing phase $\phi$.  It has recently been
proposed to use the singly Cabibbo suppressed (SCS) $D\to K^*K$ modes
to determine the mixing parameters~\cite{Xing:2007sd,Grossman:2006jg}.
However, if $\phi$ is zero, these methods would be feasible only if
the strong phase involved is measured elsewhere. The strong phase can
be measured using a Dalitz plot analysis~\cite{Rosner:2003yk}. In the
absence of the strong phase information, these modes cannot be used to
determine the mixing parameters accurately when $\phi$ is small, since
they can only be expressed as ratio of small quantities.

Our study of the various modes allows us to conclude that in the limit
of small $\phi$, an accurate measurement of all mixing parameters is
possible only if the method also allows the determination of the
parameters in the case $\phi=0$. While mixing parameters can be
determined using decays to SCS non-CP eigenstates alone, an accurate
measurement of mixing parameters in the limit of small $\phi$ is
possible, only by adding information from decays to CP-eigenstates or
if the strong phase is measured independently
elsewhere~\cite{Xing:2007sd, Grossman:2006jg, Rosner:2003yk,
  Golowich:2001hb, Gronau:2001nr,Cheng:2007uj}.  The $D\to
K^{*0}\pi^0$ modes are an example where it is possible to measure all
the mixing parameters and the strong phase using only related final
states, thereby reducing systematic errors. The methods discussed in
this letter do not have systematic errors associated with the
parameterization of the resonant content of the Dalitz
plot~\cite{Abe:2007rd} and hence are model-independent.

The neutral $D$ mass eigenstates are related to the weak eigenstates
by,  $|D_{1,2}\rangle=p|D^0\rangle\pm q|\bar{D}^0\rangle~$.
The mass and width differences of these eigenstates are popularly
written~\cite{PDG_asner} in terms of the dimensionless variables,
\[x\equiv\dsp\frac{\Delta M}{\Gamma}=\dsp\frac{M_1-M_2}{\Gamma}
\quad\text{and}\quad y\equiv\dsp\frac{\Delta
  \Gamma}{2\,\Gamma}=\dsp\frac{\Gamma_1-\Gamma_2}{2\,\Gamma},\] where
$\Gamma$ is the average of the widths of the two mass eigenstates. If
the magnitude of $q/p$ differs from unity and/or the weak phase
$\phi=\arg(q/p)$ is nonvanishing, this would signal $CP$ violation. We
consider mixing to be the only source of $CP$ violation and assume
that the decay amplitudes themselves have no weak
phase~\cite{Bergmann:2000id}.

In the limit $x\ll 1$, $y\ll 1$ and $\Gamma\,t\ll 1$, the time
dependent decay rates for a $D^0$ decaying to a final state $f$ and
$D^0\to f$ and $\bar{D}^0\to \bar{f}$ have the form:
\begin{eqnarray}
  \label{eq:ADt2barf}
  |A(D^0(t)\to\! f)|^2 \!&= e^{-\Gamma
    t}\Big[X_{\!f}+Y_{\!f} \Gamma t+Z_{\!f}(\Gamma t)^2+\!\cdots
  \Big]~\\
  |A(\bar{D}^0(t)\to\! \bar{f})|^2\! &= e^{-\Gamma t}
  \Big[\bar{X}_{\!f}+\bar{Y}_{\!f}\Gamma
  t+\bar{Z}_{\!f}(\Gamma t)^2+\! \cdots \Big]. 
\end{eqnarray}

We first consider the DCS mode $D^0\to K^{*0}\pi^0$ and its conjugate
mode $\bar{D}^0\to \bar{K}^{*0}\pi^0$, with the $K^{*0}/\bar{K}^{*0}$
reconstructed in the self-tagging $K^\pm\pi^\mp$ modes.  The
coefficient functions of the constant, linear and quadratic terms in
$(\Gamma\,t)$ in the time dependent decay rates are given by,
\begin{eqnarray}
  \label{eq:X}
{X_\kstpi} &=& \bar{X}_\kstpi =\Akstpi^2\,r_\kstpi^2 , 
\\
\label{eq:Y}
{Y_\kstpi} &=&\big|\frac{q}{p}\big| \Akstpi^2\, 
r_\kstpi \big(y^\prime_\kstpi\cos\phi-x^\prime_\kstpi\sin\phi\big),\\  
\label{eq:Ybar}
\bar{Y}_\kstpi &=& \big|\frac{p}{q}\big| \Akstpi^2\,r_\kstpi
\big(y^\prime_\kstpi \cos\phi+ x^\prime_\kstpi\sin\phi\big),  \\
\label{eq:Z}
Z_\kstpi &=& \big|\frac{q}{p}\big|^{2} \Akstpi^2\,\frac{x^2+y^2}{4} \qquad
\textrm{and}\qquad \\
\bar{Z}_\kstpi &=&  \big|\frac{p}{q}\big|^{2} \Akstpi^2\,\frac{x^2+y^2}{4},
\end{eqnarray}
where,
\begin{eqnarray}
  \label{eq:xprimekstpi}
   x^\prime_\kstpi&=&(x\cos\delta_\kstpi+y\sin\delta_\kstpi),\\
  \label{eq:yprimekstpi}
   y^\prime_\kstpi&=&(y\cos\delta_\kstpi-x\sin\delta_\kstpi),
\end{eqnarray}
with $A_{\kstpi}\equiv A(D^0\to \bar{K}^{*0}\pi^0)$ and the ratio of the
  DCS to CF amplitude defined as
\begin{equation*}
  \label{eq:rkstpi}
-r_\kstpi e^{-i\delta_\kstpi}\!\equiv\!\frac{A(D^0\to K^{*0}\pi^0)}{A(D^0\to
  \bar{K}^{*0}\pi^0 )}\!=\frac{A(D^0\to K^{*0}\pi^0)}{A(\bar{D}^0\to
  K^{*0}\pi^0 )}. 
\end{equation*}
The amplitude $\Akstpi$ can easily be measured using the time integrated
rate for
the Cabibbo favored (CF) mode $D^0\to \bar{K}^{*0}\pi^0$ which is given
by,
\begin{eqnarray}
  \label{eq:cf}
 \int_0^\infty |A(D^0(t)\to \bar{K}^{*0}\pi^0)|^2 dt &\approx& \Akstpi^2,
\end{eqnarray}
where terms of the order of $x^2$ or $y^2$ and $r_\kstpi\, x$ or
$r_\kstpi\, y$ are neglected compared to unity, as these are expected
to be ${\cal O}(10^{-4})$ or less. The ratio $r_\kstpi$ can be
determined using Eqs.~(\ref{eq:X}) and (\ref{eq:cf}).  The observables
$Z_\kstpi$ and $\bar{Z}_\kstpi$ also readily determine $|q/p|$ and
$x^2+y^2$ to be:
\begin{eqnarray}
  \label{Rm4}
  \big|\frac{q}{p}\big|^4&=&\dsp\frac{Z_\kstpi}{\bar{Z}_\kstpi},\\
  \label{eq:xsqplusysq}
  f^2\equiv x^2+y^2&=&4\dsp \frac{\sqrt{ Z_\kstpi
      \bar{Z}_\kstpi}}{\Akstpi^2}. 
\end{eqnarray}

The two linear terms in the time dependent DCS decay rates $Y_\kstpi$
and $\bar{Y}_\kstpi$ may be re-expressed in terms of two more
convenient observables $\yp{\kstpi}$ and $\ym{\kstpi}$ as follows
\begin{eqnarray}
  \label{eq:s+}
  \yp{\kstpi}=\frac{\bar{Y}_\kstpi\,|q|^2+Y_\kstpi\,|p|^2}
  {2\,r_\kstpi\,\Akstpi^2|q|\,|p|} &=& y^\prime_\kstpi \cos\phi,\\  
  \label{eq:s-}
  \ym{\kstpi}=\frac{\bar{Y}_\kstpi\,|q|^2-Y_\kstpi\,|p|^2}
  {2\,r_\kstpi\,\Akstpi^2|q|\,|p|} &=& x^\prime_\kstpi\sin\phi. 
\end{eqnarray}
Note that the observable $\ym{\kstpi}$ may be difficult to measure in
the small $\phi$ limit.

The $K^{*0}/\bar{K}^{*0}$ in the final state could also have been
reconstructed in the neutral $K_{\sss\! S}\pi^0$ mode, resulting in an
additional observable. A unique feature of the final state
$K_{\sss\!S}\pi^0\pi^0$ is that it includes contributions from both
$K^{*0}\pi^0$ as well as $\bar{K}^{*0}\pi^0$ states; the amplitude for
this final state is thus a sum of the CF and DCS amplitudes,
\begin{eqnarray}
  \label{eq:kspipi}
\Akpipi^2 &\equiv& |A(D^0\to K_{\sss\! S}\pi^0\pi^0)|^2\nn  \\
&=&|A_\kstpi|^2(1+r_\kstpi^2-2\,r_\kstpi\cos\delta_\kstpi). 
\end{eqnarray}
Since the decay mode involves two neutral pions it will not be easy to
perform a time dependent study. Hence, we consider only the time
integrated decay rate for this mode.  The amplitudes $A(D^0\to
K_{\sss\! S}\pi^0\pi^0)$ and $A(\bar{D}^0\to K_{\sss\! S}\pi^0\pi^0)$
are equal since $K_{\sss\! S}\pi^0\pi^0$ is a CP eigenstate. Hence,
the time integrated decay rate for $D^0\to K_{\sss\! S}\pi^0\pi^0$ is
given by:
\begin{eqnarray}
\lefteqn{\int_0^\infty |A(D^0(t)\to K_{\sss\! S}\pi^0\pi^0)|^2 dt}\nn\\
\!&\!\!&\!\!\approx\!
\Akpipi^2\!\big[1\!+\!\frac{q}{p}\!(y\,\cos\phi-x\,\sin\phi)\big]\nn\\ 
  \label{eq:ADt2f}
\!&\!\!&\!\!\approx\! |A_\kstpi|^2\!\big[1\!+\!\frac{q}{p}\!(y\,\cos\phi-x\,\sin\phi)\!
-\!2 r_{\kstpi}\cos\delta_\kstpi\big],
\end{eqnarray}
where, terms of order $x^2$, $y^2$ and $r_\kstpi\,x$, $r_\kstpi\,y$ as
well as $r_\kstpi^2$ are once again neglected compared to unity.

Using Eqs.~(\ref{eq:s+}) and (\ref{eq:s-}), one obtains the following
solutions for $\tan^2\phi$ and $y^{\prime 2}_\kstpi$:
\begin{eqnarray}
  \label{eq:cphi}
  \tan^2\phi\!&=&\!\frac{2f^2-{\cal F_\kstpi}-
    \sqrt{{\cal F_\kstpi}^2-4f^2\yp{\kstpi}^2}}{{\cal F_\kstpi}+
    \sqrt{{\cal F_\kstpi}^2-4f^2\yp{\kstpi}^2}}  \\  
  y^{\prime 2}_\kstpi\!&=&\!\frac{{\cal F_\kstpi}-
    \sqrt{{\cal F_\kstpi}^2-4f^2
      \yp{\kstpi}^2}}{2}
\end{eqnarray}
where, ${\cal F}_\kstpi=f^2-\ym{\kstpi}^2+\yp{\kstpi}^2$. The ambiguity
in the solutions of the quadratic equations in $\tan^2\phi$ and
$y^{\prime 2}_\kstpi$ is fixed by the correct limiting solution in the
$\phi=0$ limit. Further, expressing $\cos\delta_\kstpi$ in terms of
$x$, $y$ and $x^\prime_\kstpi$, $y^\prime_\kstpi$,
Eq.~(\ref{eq:ADt2f}) may be rewritten as a quadratic equation in
$x/y$,
\begin{equation}
  \label{eq:xbyy}
  \big[B^2-\zeta^2\big]\,\Big(\frac{x}{y}\Big)^2+
  2\,A\,B\,\frac{x}{y}+A^2-\zeta^2=0,  
\end{equation}
where,
\begin{gather}
  \label{eq:coeff}
  A=2
  r_\kstpi\,{y^\prime_\kstpi}-\Big|\frac{q}{p}\Big|\frac{\yp{\kstpi}
    f^2}{y^\prime_\kstpi},\quad
  B=2\,r_\kstpi\,{x^\prime_\kstpi}+\Big|\frac{q}{p}\Big|\frac{\yp{\kstpi}
    f^2}{x^\prime_\kstpi},\nn\\ 
 \zeta= \Bigg(\frac{Br(D^0\to
   K_{\sss\! S}\pi^0\pi^0)}{Br(D^0\to \bar{K}^{*0}\pi^0)}\,-\,1\Bigg)\,f,
\end{gather}
allowing $x/y$ to be solved with a four-fold ambiguity. $x$ and $y$
can thus be individually determined using Eq.~(\ref{eq:xsqplusysq}).
The solution obtained is finite even if $\phi=0$, with a correction
term of order $\ym{\kstpi}$. Hence an accurate estimation is possible
even if $\phi$ is tiny.  We show below that the ambiguity in $x/y$ can
be reduced if information from $K^+K^-$ modes is added as well.

We next consider the time dependent decay of a D meson to a singly
Cabibbo suppressed (SCS) CP eigenstate such as $D\to K^+K^-$ or
$D\to\pi^+\pi^-$.  To be specific, we will consider only the $K^+K^-$
final state, but the conclusions can be straightforwardly applied to
any other SCS-CP eigenstate. For this final state, the strong phase is
identically zero; and hence, the coefficients of the constant and
linear terms in $(\Gamma\,t)$, defined using the time dependent decay
in Eq.(\ref{eq:ADt2barf}) reduce to the simple form:
\begin{eqnarray}
  \label{eq:XCP}
  {X}_\kk &=& \bar{X}_\kk =\big|A_\kk\big|^2\,\\
  \label{eq:YCP}
  {Y}_\kk &=& -\big|\frac{q}{p}\big| \big|A_\kk\big|^2 (-x \sin\phi +
  y\cos\phi),\\ 
  \label{eq:YbarCP}
  \bar{Y}_\kk &=&  -\big|\frac{p}{q}\big| \big|A_\kk\big|^2 (x
  \sin\phi + y\cos\phi)   
\end{eqnarray}
Unlike the DCS modes where the term quadratic in $\Gamma\,t$ is
enhanced by the ratio of CF to DCS rates, in the SCS modes all time
dependent terms are of the same order in $\sin\theta_c$, hence
quadratic and higher terms in $\Gamma\,t$ cannot be extracted.
Assuming $|q/p|\approx 1$ and $\phi=0$, the linear term in $\Gamma\,t$
can directly measure $y$ as has been done in
Ref.~\cite{Staric:2007dt}.  However, the time dependent study of only
the SCS CP eigenstates does not allow $x$ to be determined, even in
the limit $|q/p|\approx 1$ and $\phi=0$.

We will show that if we also include in this analysis the quadratic
terms in $(\Gamma\,t)$ from the time dependent decay rates of DCS
modes such as $K\pi$, all the mixing parameters can be solved without
approximation.  For $D^0\to K^+\pi^-$ and $\bar{D}^0\to K^-\pi^+$, the
coefficient functions of the quadratic terms in $(\Gamma\,t)$ will be
analogous to those for the $K^*\pi$ mode given in Eq.~(\ref{eq:Z}).
Hence, the corresponding observables $Z_\kpi$ and $\bar{Z}_\kpi$
readily determine $|q/p|$ and $f^2=x^2+y^2$. Alternatively, $|q/p|$
and $f^2$ could be measured using time integrated wrong sign relative
to right sign semileptonic decay rates. Having obtained $|q/p|$ and
$f^2$, $\phi$ and $x/y$ can easily be determined from $D\to K^+K^-$.
Using Eqs.~(\ref{eq:XCP}) -- (\ref{eq:YbarCP}), which can be
re-expressed as,
\begin{eqnarray}
  \label{eq:ycosphi}
  \yp{\kk}&=&\frac{\bar{Y}_\kk\,|q|^2 +
    Y_\kk\,|p|^2}{2\,X_\kpi|q|\,|p|} = -y\cos\phi,\\ 
  \label{eq:xsinphi} 
  \ym{\kk}&=&\frac{\bar{Y}_\kk\,|q|^2 -
    Y_\kk\,|p|^2}{2\,X_\kpi|q|\,|p|}  = -x\sin\phi, 
\end{eqnarray}
solution for $x^2/y^2$ and $\phi$ can be straightforwardly written
\begin{gather*}
  \label{eq:ysq}
  \frac{x^2}{y^2}=\frac{{\cal F}_\kk-2\yp{\kk}^2+
    \sqrt{{\cal F}_\kk^2
      -4\,f^2\,\yp{\kk}^2}}{2\yp{\kk}^2},\\
\tan^2\phi=\frac{2f^2-{\cal F}_\kk-
    \sqrt{{\cal F}_\kk^2 -4\,f^2\,\yp{\kk}^2}}{{\cal F}_\kk+
    \sqrt{{\cal F}_\kk^2 -4\,f^2\,\yp{\kk}^2}},
\end{gather*}
where ${\cal F}_\kk=f^2+\yp{\kk}^2-\ym{\kk}^2$.  We once again examine
in detail the solution for the case of small $\phi$. If $\phi$ is
small, the measured value of $\ym{\kk}$ will be small. The above  solutions
can then be written as a series in $\ym{\kk}^2$:
\begin{gather*}
  \label{eq:ysqbyxsq}
 \frac{x^2}{y^2}=\frac{f^2-\yp{\kk}^2}{\yp{\kk}^2}-
 \frac{\ym{\kk}^2\,f^2}{\yp{\kk}^2(f^2-\yp{\kk}^2)} + {\cal O}(\ym{\kk}^4) \\
\label{eq:cossqphi}
\tan^2\phi=\frac{\ym{\kk}^2}{f^2-\yp{\kk}^2}+ {\cal O}(\ym{\kk}^4),
\end{gather*}
and therefore $x^2/y^2$ is finite even for small $\phi$.

As mentioned earlier, if information from $K^+K^-$ modes is added to that
from the $K^*\pi$ modes, a reduction in ambiguity is possible.  If the
observables $\yp{\kk}$ and $\ym{\kk}$ are used, then
$\cos\delta_\kstpi$ can be obtained purely in terms of observables
directly from Eq.~(\ref{eq:ADt2f}). Knowing $\cos\delta_\kstpi$,
Eqs.~(\ref{eq:s+}), (\ref{eq:s-}) and (\ref{eq:ycosphi}),
(\ref{eq:xsinphi}) can be used to get,
\begin{equation*}
  \frac{x^2}{y^2}=\frac{\big(\dsp \yp{\kstpi}-
    \cos\delta_\kstpi\yp{\kk}\big)^2}{\yp{\kk}^2(1-\cos^2\delta_\kstpi)}.  
\end{equation*}
Combining this with the sum $x^2+y^2$, $x^2$ and $y^2$ can be
individually determined. Further, Eqs.(\ref{eq:xsinphi}) and (\ref{eq:s-})
can be used to obtain,
\begin{equation}
  \label{eq:sd}
  \frac{y}{x}=\frac{-1}{\sin\delta_\kstpi}\Big(\frac{\ym{\kstpi}}{\ym{\kk}}+
  \cos\delta_\kstpi\Big),
\end{equation}
which helps in reducing the ambiguities in $x$ and $y$ from four-fold to
two-fold.

Recently a method was proposed~\cite{Xing:2007sd} to determine all the
mixing parameters using the $D\to K^*K$ modes.  This mode is singly
Cabibbo suppressed (SCS) but unlike the $K^+K^-$ mode it is not a CP
eigenstate. Hence one can study time dependence in all the four modes:
$D^0\to K^{*\pm}K^\mp$ and $\bar{D}^0\to K^{*\mp}K^\pm$. The
coefficients of the constant and linear terms in $(\Gamma\,t)$ may be
written as:
\begin{eqnarray}
  \label{eq:Ykst+k-}
  Y_\kstk\!&\!=\!&\!\big|\frac{q}{p}\big|\Akstkno^2
  r_\kstkno\Big(y^\prime_\kstk 
  \cos\phi-x^\prime_\kstk \sin\phi\Big)\nn\\ 
  \label{eq:Ybarkst+k-}
  \bar{Y}_\kstk\!&\!=\!&\!\big|\frac{p}{q}\big|\Akstkno^2
  r_\kstkno\Big(y^\prime_\kstk 
  \cos\phi+x^\prime_\kstk 
  \sin\phi\Big)\nn\\ 
  \label{eq:Ykst-k+}
  Y_\kstkbar\!&\!=\!&\!\big|\frac{q}{p}\big|\Akstkno^2
  r_\kstkno\Big(y^\prime_\kstkbar 
  \cos\phi-x^\prime_\kstkbar 
  \sin\phi\Big)\nn\\
  \label{eq:Ybarkst-k+}
  \bar{Y}_\kstkbar\!&\!=\!&\!\big|\frac{p}{q}\big|\Akstkno^2
  r_\kstkno\Big(y^\prime_\kstkbar \cos\phi+x^\prime_\kstkbar 
  \sin\phi\Big)\nn\\
  \label{eq:Xkst+k-}
  X_\kstk\!&\!=\!&\!\bar{X}_\kstk=\Akstkno^2\nn\\
  \label{eq:Xbarkst-k+}
  X_\kstkbar\!&\!=\!&\!\bar{X}_\kstkbar=\Akstkno^2 r_\kstkno^2
\end{eqnarray}
where, $A_\kstkno=A(D^0\to K^{*+}K^-)=A(\bar{D}^0\to K^{*-}K^+)$ and
$r_\kstkno$ is defined as
\begin{equation*}
  -r_\kstkno\,e^{i\delta_\kstkno}=\frac{A(\bar{D}^0\to
    K^{*+}K^-)}{A(D^0\to K^{*+}K^-)}=\frac{A(D^0\to 
    K^{*-}K^+)}{A(\bar{D}^0\to  K^{*-}K^+)}.
\end{equation*}
It may also be noted that $y^\prime_\kstkbar$ and $x^\prime_\kstkbar$
are different from $y^\prime_\kstk$ and $x^\prime_\kstk$ and are
defined as:
\begin{eqnarray}
  \label{eq:xprimekstk}
   x^\prime_{\kstk,\kstkbar}&=&(x\cos\delta_\kstkno\pm
   y\sin\delta_\kstkno),\nn\\ 
 \label{eq:yprimekstk}
   y^\prime_{\kstk,\kstkbar}&=&(y\cos\delta_\kstkno\mp
   x\sin\delta_\kstkno).
\end{eqnarray}
One may conclude that the six observables in Eqs.~(\ref{eq:Xkst+k-})
can be used to evaluate the six parameters $\Akstkno^2$,
$r_\kstkno^2$, $x$, $y$, $\phi$ and $\delta_\kstkno$ assuming the
value of $|q/p|$ from elsewhere. However, note that if the mixing
phase $\phi=0$, then, the number of observables reduces to four (since
now, $|q|^2\bar{Y}_\kstk=|p|^2Y_\kstk$ and
$|q|^2\bar{Y}_\kstkbar=|p|^2Y_\kstkbar$) and a solution of all the
five parameters is not possible without some additional information.
Moreover, for small but nonvanishing $\phi$ the solution for the ratio
$x^2/y^2$ will be inaccurate as it will depend on ratio of two very
small observables. To see this, let us define,
$\yp{\kstk}=y^\prime_\kstk \cos\phi$, $\ym{\kstk}= x^\prime_\kstk
\sin\phi$, $\yp{\kstkbar}=y^\prime_\kstkbar \cos\phi$ and
$\ym{\kstkbar}=x^\prime_\kstkbar \sin\phi$, which can all be 
determined in terms of observables using Eqs.~(\ref{eq:Xkst+k-}).
It then, it follows that:
\begin{eqnarray}
  \label{eq:xbykstk}
  \frac{x^2}{y^2}&\!=\!&\frac{(\ym{\kstk}+\ym{\kstkbar})
    (\yp{\kstkbar}-\yp{\kstk})}
  {(\ym{\kstk}-\ym{\kstkbar}) (\yp{\kstkbar}+\yp{\kstk})}. 
\end{eqnarray}
It is clear that the RHS involves the ratio of two small quantities
when $\phi$ is small. It is easy to see that this situation is easily
alleviated if $\delta_\kstkno$ is measured
elsewhere~\cite{Rosner:2003yk}. In fact, the knowledge of
$\delta_\kstkno$ not only allows the additional determination of
$|q/p|$~\cite{Grossman:2006jg}, but also enables an accurate
measurement of mixing parameters. 

We now estimate the values of the mixing parameters that can be
obtained using the current data for $D\to
K^+K^-/\pi^+\pi^-$~\cite{Staric:2007dt} and the world average for
$x^2+y^2$~\cite{HFAG}. Assuming $|q/p|=1$, we obtain $\yp{\kk}=0.0131
\pm 0.0041$, $\ym{\kk}= -0.0001 \pm 0.0034$ and $f^2=0.00042 \pm
0.00022$, resulting in $|x|=(1.57\pm 0.56)\times 10^{-2}$,
$|y|=(1.31\pm 0.41)\times 10^{-2}$ and value (up to ambiguities) of
$\phi =\pm(0.36 \pm 12.36)^o$. We emphasize that our method allows the
determination of mixing parameters even for $|q/p|\neq 1$; the choice
$|q/p|=1$ has been made here, only due to lack of complete tabulated
data.

An estimate of the precision to which the mixing parameters can be
measured, using the $D\to K^{*}\pi$ modes, requires the number of
reconstructed $D\to K^{*0}\pi^0\to K^+\pi^-\pi^0$ events. While a
branching fraction for this mode has not yet been reported, about 500
events (in $230 fb^{-1}$) for the mode $D^0\to K^{*+}\pi^-\to
K^+\pi^0\pi^-$, have been observed~\cite{Aubert:2006kt}.  We present
our estimates using two representative values for the ratio of DCS
modes:
\begin{equation*}
  \frac{B(D^0\to K^{*0}\pi^0\to K^+\pi^-\pi^0)}{B(D^0\to K^{*+}\pi^-\to
    K^+\pi^0\pi^-)}=(0.4,1.2).
\end{equation*} 
These values are chosen to be of the order $0.85$, the
measured~\cite{Yao:2006px} ratio of corresponding CF branching
fractions. With an integrated luminosity of $1\,ab^{-1}$ at an $e^+
e^-$ B factory we expect about $(4000,12000)$ $D^0\to K^{*0}\pi^0\to
K^+\pi^-\pi^0$ events. Interpolating the errors in $D\to K^+\pi^-$ and
assuming $\delta=\phi=0$, the approximate errors on $|x|^2$ and $|y|$
are expected to be $(4.7,2.7)\times 10^{-4}$ and $(8.9,5.2)\times
10^{-3}$, respectively.

We have proposed a new method to determine the $D^0-\bar{D}^0$ mixing
parameters $x$, $y$, $|q/p|$ and $\phi$ for arbitrary values of
$\phi$.  The doubly Cabibbo suppressed mode $D^0\to K^{*0} \pi^0$
reconstructed in two final states ($K^+ \pi^- \pi^0$ and $K_{\sss\! S}
\pi^0 \pi^0$) enables the determination of all the mixing parameters.
For the $K_{\sss\! S} \pi^0 \pi^0$ mode, only time integrated
measurements are used, while for the $K^+\pi^-\pi^0$ mode time
dependent measurements are required.  We also show that decays to the
CP eigenstate $D\to K^+K^-$ together with $D\to K^+\pi^-$ can be used
to extract all the mixing parameters. By combining measurements of $D
\to K^{*0} \pi^0$ with results on $D\to K^+ K^-$ one can reduce the
number of ambiguous solutions for mixing 
parameters. We estimate that $|x|$, $|y|$ and $\phi$ can be measured
with precision of order $0.6\times 10^{-2}$, $0.4\times 10^{-2}$ and
$12^o$ respectively, using data available at present. It should be
possible to determine $|x|$, $|y|$ to order $7\times 10^{-4}$,
$4\times 10^{-4}$ respectively and $\phi$ to about $1^o$ 
at a Super-B factory with an integrated luminosity of
$50\,ab^{-1}$~\cite{Super-B}.

The work of N.G.D was supported in part by the US DOE under Grant No.
DE-FAG02-96ER40969. T.E.B. and S.P. were supported by the US DOE under
Contract DE-FG02-04ER41291. N.S. was supported in part by DST, India.

\end{document}